\documentclass[a4paper,11pt]{article}
\usepackage{pos}

\usepackage{slashed}
\global\long\def\not#1{\slashed{#1}}%

\title{On U-Folds and Their Construction}
\author[b,c]{Stefano Maurelli}
\author[a,b]{Ruggero Noris}
\author[b,d]{Marcelo Oyarzo}
\author*[b,c]{Mario Trigiante}
\affiliation[a]{CEICO, Institute of Physics of the Czech Academy of Sciences,
Na Slovance 2, 182 21 Prague 8, Czech Republic}
\affiliation[b]{Department of Applied Science and Technology, Politecnico di Torino, Corso Duca degli Abruzzi, 24, 10129 Torino, Italy}
\affiliation[c]{INFN, Sezione di Torino, Via P. Giuria 1, 10125 Torino, Italy}
\affiliation[d]{Departamento de F\'isica, Universidad de Concepci\'on Casilla, 160-C, Concepci\'on, Chile.}
\emailAdd{ruggeronoris28@gmail.com}
\emailAdd{stefano.maurelli@polito.it}
\emailAdd{moyarzoca1@gmail.com}
\emailAdd{mario.trigiante@polito.it}
\abstract{We review a general paradigm for constructing U-fold backgrounds in (dimensionally reduced) Type IIB superstring theory, of the form ${\rm AdS}_{d-1}\times S^1\times S^d$, with a monodromy along $S^1$ in the string-duality group. We also consider a special instance with $d=3$ in Type IIB superstring theory, discuss its ten-dimensional uplift and assess its supersymmetry.}

\FullConference{Proceedings of the Corfu Summer Institute 2024 "School and Workshops on Elementary Particle Physics and Gravity" (CORFU2024)\
12 - 26 May, and 25 August - 27 September, 2024\\
Corfu, Greece\\}

\begin{document}
\maketitle

\section{Introduction}
Superstring theory is widely acknowledged to be a promising candidate to provide a unifying quantum framework where to describe the fundamental interactions.
However, while the latter quantum theory is expected to be unique, there are five superstring theories.
This seeming contradiction was solved in the '90s through the discovery of \emph{superstring dualities} \cite{Polchinski:1998rr}, which are going to play a central role in our discussion.
Superstring dualities are correspondences between superstring theories on different backgrounds, in light of which these theories can be regarded as distinct, effective descriptions of the same quantum physics. Part of such dualities, as mappings between string backgrounds, manifest themselves as (discrete) global symmetries in the low-energy supergravity description, described by a group $G(\mathbb{Z})$. For instance, the quantum global symmetry group ${\rm E}_{7(7)}(\mathbb{Z})$ of four-dimensional maximal supergravity contains the (proper) T-duality group ${\rm SO}(6,6;\mathbb{Z})$ and the Type IIB S-duality ${\rm SL}(2;\mathbb{Z})$. This led the authors of  \cite{Hull:1994ys} to conjecture that ${\rm E}_{7(7)}(\mathbb{Z})$ be an exact symmetry of the yet unknown unifying quantum theory underlying Type II superstrings (\emph{U-duality}).
This makes supergravity a privileged framework for studying string dualities, and ungauged supergravities have indeed provided valuable test grounds for these correspondences.\par
On the other hand, gauged extended supergravities in $d$-dimensions have provided a valuable framework to construct and study solutions to ten or eleven-dimensional supergravities, whose spacetime geometries have the general form of a warped product ${\rm  M_d}\times_{{\rm w}} {\rm M_{int.}}$ of a $d$-dimensional, non-compact spacetime and an internal compact manifold ${\rm M_{int.}}$. These  $d$-dimensional models describe a \emph{consistent truncation} of the low-lying modes of the ten or eleven-dimensional parent theory on such background and feature ${\rm M_d}$ as a solution. When ${\rm M_d}$ is maximally symmetric (e.g. a $d$-dimensional anti-de Sitter spacetime ${\rm AdS}_d$), it defines a vacuum of the $d$-dimensional truncation, identified by an extremum $\phi_0$ of the scalar potential $V(\phi)$. The geometry of ${\rm M_d}$, as well as any other constant background quantity like form-fluxes, are encoded in a characteristic gauge group $\mathcal{G}$ of the $d$-dimensional supergravity and in $\phi_0$. This group and the extended supersymmetry of the model uniquely fix the scalar potential $V(\phi)$ as well as the mass spectrum of the truncation at $\phi_0$. This lower-dimensional supergravity approach is particularly useful for constructing ten or eleven-dimensional backgrounds
with low residual symmetries since the problem of solving differential equations in the higher-dimensional fields is reduced to the algebraic task of extremizing a scalar potential. Although not all backgrounds of the aforementioned form admit such a description, the \emph{embedding tensor formulation} of gauged extended supergravities (see \cite{Samtleben:2008pe} and \cite{Trigiante:2016mnt} for reviews) and the discovery of the so-called \emph{dyonic gaugings} \cite{DallAgata:2011aa,DallAgata:2012mfj} (see also \cite{Inverso:2025zct} for a recent update) have allowed the construction of lower-dimensional supergravity descriptions of several new compactifications of superstring or eleven-dimensional supergravity. Moreover \emph{Exceptional Field Theory} (ExFT) \cite{Hohm:2013pua,Hohm:2013uia,Hohm:2014qga,Samtleben:2025fta} and \emph{Exceptional Generalized Geometry} (EGG) \cite{Coimbra:2011ky,Coimbra:2012af,Lee:2014mla,Cassani:2019vcl} have provided a direct embedding of certain gauged lower-dimensional maximal models into Type II or eleven-dimensional supergravities.\par
In this contribution, we shall focus on a particular class of supergravity solutions, belonging to the general class of $U$-folds \cite{Hull:2003kr,Hull:2004in} (see also \cite{Plauschinn:2018wbo} for a recent review), which encode superstring dualities as a built-in feature: as we move in space, the background fields describing them undergo a duality transformation. Such solutions are globally non-geometric in that they require more than one spacetime coordinate patch to be described and, on intersecting patches, the transition functions involve string dualities.\footnote{In globally geometric solutions, the transition functions only implement the effect of spacetime diffeomorphisms and gauge transformations.}
The first instance of the class of backgrounds under consideration is half-supersymmetric (i.e. preserves 16 supercharges) and  was constructed by uplifting to Type IIB superstring theory \cite{Inverso:2016eet} the unique
$\mathcal{N}=4$ vacuum \cite{Gallerati:2014xra} of the maximal $D=4$ supergravity with gauge group $\mathcal{G}=[{\rm SO}(6)\times {\rm {\rm SO}(1,1) }]\ltimes \mathbb{R}^{12}$.
The geometry of the ten-dimensional solution is ${\rm AdS}_4\times S^1\times \tilde{S}^5$, where $\tilde{S}^5$ has the topology of a 5-sphere and there is a
monodromy $\mathfrak{M}$ along $S^1$, described by a hyperbolic element of the ten-dimensional S-duality group of Type IIB superstring theory: $\mathfrak{M}=J_{\tt n}\equiv \left(\begin{matrix}{\tt n} & 1\cr -1 &0\end{matrix}\right)\in {\rm SL}(2,\mathbb{Z})_{{\rm IIB}}$,  ${\tt n}>2$. This solution is also referred to as $J$-fold and can be obtained through a suitable compactification of a Janus solution ${\rm AdS}_4\times \mathbb{R}\times \tilde{S}^5$ \cite{DHoker:2007hhe,DHoker:2007zhm}. This suggests a holographically dual description in terms of a strongly coupled $D=3$ SCFT defined on an interface in a $D=4$ ${\rm U}(N)$ SYM theory, compactified on an $S^1$ orthogonal to the interface, with a monodromy $J_{\tt n}$ along the circle acting on the complexified coupling constant. A precise definition of this dual $D=3$, $\mathcal{N}=4$ SCFT theory  ($J$-fold theory) was put forward in \cite{Assel:2018vtq} as the IR-limit of a ${\rm T[U(N)]}$ model \cite{Gaiotto:2008ak} in which the diagonal subgroup of the ${\rm U}(N)\times {\rm U}(N)$ global symmetry is gauged by the $\mathcal{N}=4$ vector multiplet and a level-${\tt n}$ Chern-Simons term is added. The $\mathcal{N}=4$ vacuum is part of a three-parameter class of extrema of the scalar potential $V(\phi)$, all connected by flat directions which are holographically dual to exactly marginal deformations of the main $J$-fold model and which may either preserve eight of the 16 supercharges or none. All these vacua uplift to deformations of the half-maximal $J$-fold. Other families of marginally connected vacua were found in the same $\mathcal{N}=8$ gauged supergravity in \cite{Guarino:2019oct}, and uplifted to J-folds. There has been extensive research work on this class of Type IIB solutions, variants thereof, and their dual description
\cite{Bobev:2019jbi,Guarino:2020gfe,Bobev:2020fon,Giambrone:2021zvp,Guarino:2021kyp,Arav:2021gra,Bobev:2021yya,Guarino:2021hrc,Cesaro:2021tna,Giambrone:2021wsm,Guarino:2022tlw,Bobev:2024mqw,Guarino:2024zgq,Guarino:2024gke}.\par
In all these solutions, the dependence of the fields on the coordinate $\eta$ of $S^1$, $\eta\in [0,T]$, is encoded in a twist matrix $\mathcal{A}(\eta)$, contained in the global symmetry group ${\rm SL}(2,\mathbb{R})_{{\rm IIB}}$ of classical Type IIB supergravity, and acting on the Type IIB fields according to their transformation property under this group. The resulting monodromy matrix $\mathfrak{M}$ is $\mathcal{A}(\eta)^{-1}\cdot \mathcal{A}(\eta+T)$. In particular, by the effect of $\mathcal{A}(\eta)$, the axion-dilaton field, as we move around the circle, spans a \emph{geodesic} in the moduli space. This suggests that the $\mathcal{N}=8$ gauged supergravity under consideration, whose vacua uplift to $J$-folds, can be obtained as a Cremmer-Scherk-Schwarz (CSS) reduction \cite{Cremmer:1979uq} of the ${\rm SO}(6)$-gauged maximal five-dimensional model on a circle, with hyperbolic CSS twist $\mathcal{A}(\eta)$ in the global symmetry group ${\rm SL}(2,\mathbb{R})_{{\rm IIB}}$ of the theory \cite{Inverso:2016eet}. The simplest $J$-fold solution is non-supersymmetric and has symmetry ${\rm SO}(6)$ \cite{Guarino:2019oct}. Its spacetime geometry is ${\rm AdS}_4\times S^1\times S^5$ and features no 2-form fields. This solution suggests a general procedure for constructing ${\rm AdS}$ $U$-folds in Type IIB supergravity \cite{Astesiano:2024gzy}:
\begin{itemize}
    \item{Consider a solution of the form ${\rm AdS}_d\times S^d\times M_{10-2d}$, $d$ odd, with moduli fields $\varphi=(\varphi^a)$. Here $M_{10-2d}$ is a trivial manifold for $d=5$, while it is a Calabi-Yau 2-fold ($M_{4}=CY_2$) for $d=3$;}
    \item{Compactify one direction in the boundary of ${\rm AdS}_d$;}
    \item{Give the moduli fields $\varphi$ a dependence on the coordinate $\eta$ along the compact direction on $\partial{\rm AdS}_d$, which defines a geodesic on the moduli space;}
    \item{The backreaction of the evolving moduli fields on spacetime yields a background of the form:$${\rm AdS}_{d-1}\times S^1\times S^d\times M_{10-2d}\,,$$
    with monodromy along $S^1$.}
    \item{The monodromy $\mathfrak{M}$ is defined by the global symmetry element connecting the two endpoints of the geodesic.}
\end{itemize}
In the remainder of this contribution, we shall enter into the details of this construction.
\section{General Contruction of $U$-Folds}
Consider a bosonic model in $2d$-dimensions, $d$ odd, describing Einstein gravity coupled to $n$ self-dual and $m$ anti-self-dual $(d-1)$-forms $B_{(d-1)}^M\equiv B^M_{\hat{\mu}_1\dots \hat{\mu}_{d-1}}\,dx^{\hat{\mu}_1}\wedge \dots dx^{\hat{\mu}_{d-1}}/(d-1)!$, $M=1,\dots, n+m$, $\hat{\mu},\hat{\nu}\,\dots=0,\dots, 2d-1$, and scalar fields $\phi=(\phi^s)$ described by a non-linear sigma-model with symmetric target space:
\begin{equation}
    \phi\in \mathscr{M}_{{\rm scal}}=\frac{G}{H}\,.
\end{equation}
Examples of such a model are the bosonic sectors of $\mathcal{N}=(2,0),\,(1,1)$ or $(2,2)$ six-dimensional supergravity in which the vector fields, where they exist, are truncated out. {Another example is } Type IIB supergravity in $D=10$ in which the 2-form fields are set to zero. The action of the isometry group $G$ of the scalar manifold on the scalar fields is described by the known equation of coset-geometry:
$$g\cdot L(\phi)=L(\phi')\cdot h(g,\phi)\,,$$
where $L(\phi)\in G$ is the coset representative.
As is the case of the supergravity models listed above, we assume $G$ to have a pseudo-orthogonal representation $\mathscr{R}$ defining its action on the $n$ self-dual and $m$ anti-self-dual $d$-form field strengths $H_{(d)}^M=dB_{(d-1)}^M$:\footnote{This representation may be non-faithful.}
\begin{align}
    \forall g\in G\,&\stackrel{\mathscr{R}}{\longrightarrow}\,\mathscr{R}[g]=(\mathscr{R}[g]^M{}_N)\in {\rm O}(m,n)\,\,:\,\,\,\,\mathscr{R}[g]^T\,\Omega\,\mathscr{R}[g]=\Omega\,,\nonumber\\
    \Omega&={\rm diag}(\underbrace{+,\dots,+}_m,\underbrace{-,\dots,-}_n)\,.
\end{align}
The non-minimal couplings of the scalar fields to the tensors are encoded in the (positive definite) kinetic matrix $\boldsymbol{\mathcal{M}}(\phi)=(\mathcal{M}_{MN}(\phi))$ of the matter fields, given by:
$$\boldsymbol{\mathcal{M}}(\phi)\equiv \mathscr{R}[L(\phi)]\cdot \mathscr{R}[L(\phi)]^T\in {\rm O}(m,n)\,,$$
so that, if $g\in G$ maps $\phi$ into $\phi'(\phi)$, $\mathcal{M}(\phi)$ transforms as follows:
\footnote{We use the "mostly plus" convention for the metric and define ${}^*H_{\hat{\mu}_1\dots \hat{\mu}_d}\equiv\frac{\sqrt{|{\rm det}(g)|}}{d!}\epsilon_{\hat{\mu}_1\dots \hat{\mu}_d\hat{\nu}_1\dots \hat{\nu}_d}\,H^{\hat{\nu}_1\dots \hat{\nu}_d}$ and $\epsilon_{0,\dots, 2d}=1$.}
$$\forall g\in G\,:\,\,\mathcal{M}(\phi)\stackrel{g}{\longrightarrow}\,\mathcal{M}(\phi')=\mathscr{R}[g]\cdot\mathcal{M}(\phi)\cdot \mathscr{R}[g]^T\,.$$
The field equations are written in a $G$-invariant form:
\begin{itemize}
    \item{Tensors:
    $d{\bf H}_{(d)}=0\,,\,\,{}^*{\bf H}_{(d)}=-\Omega\cdot \boldsymbol{\mathcal{M}}(\phi)\cdot {\bf H}_{(d)}$\,,}
  \item{Scalars: $D_{\hat{\mu}}(\partial^{\hat{\mu}}\phi^s)\equiv \nabla_{\hat{\mu}}\partial^{\hat{\mu}}\phi^s+\tilde{\Gamma}_{rt}\,\partial_{\hat{\mu}}\phi^r\,\partial^{\hat{\mu}}\phi^t=\mathscr{G}(\phi)^{st}\,{\bf H}^T_{\hat{\mu_1}\dots \hat{\mu_1}}\cdot \frac{\partial}{\partial \phi^t}\boldsymbol{\mathcal{M}}\cdot {\bf H}^{\hat{\mu_1}\dots \hat{\mu_1}}\,,$}
  \item{Einstein: $R_{\hat{\mu}\hat{\nu}}-\frac{1}{2}\,g_{\hat{\mu}\hat{\nu}}\,R=T_{\hat{\mu}\hat{\nu}}^{(s)}+T_{\hat{\mu}\hat{\nu}}^{(H)}$\,,}
\end{itemize}
where we have suppressed the pseudo-orthogonal indices by defining ${\bf H}_{(d)}\equiv ({ H}^M_{(d)})$, $\mathscr{G}_{rs}(\phi)$ is the Riemannian metric on the scalar manifold and $\tilde{\Gamma}$ the corresponding Levi-Civita symbol. The energy-momentum tensors read:
\begin{equation}
    T_{\hat{\mu}\hat{\nu}}^{(s)}\equiv \frac{1}{2}\,\mathscr{G}_{rs}\left(\partial_{\hat{\mu}}\phi^r\,\partial_{\hat{\nu}}\phi^s-\frac{1}{2}\,g_{\hat{\mu}\hat{\nu}}\,\partial_{\hat{\rho}}\phi^r\,\partial^{\hat{\rho}}\phi^s\right)\,,\,\,T_{\hat{\mu}\hat{\nu}}^{(H)}\equiv \frac{1}{2(d-1)!}\,{\bf H}^T_{\hat{\mu}\hat{\mu}_1\dots \hat{\mu}_{d-1}}\cdot \boldsymbol{\mathcal{M}}\cdot {\bf H}_{\hat{\nu}}{}^{\hat{\mu}_1\dots \hat{\mu}_{d-1}}\,.
\end{equation}
The above equations are $G$-invariant provided ${\bf H}_{(d)}$ transforms as follows:
$$\forall g\in G\,:\,\,{\bf H}_{(d)}\stackrel{g}{\longrightarrow}\,{\bf H}_{(d)}'=(\mathscr{R}[g]^{-1})^T\cdot{\bf H}_{(d)}\,.$$
\subsection{The solution} We look for a solution in which spacetime has the general form ${\bf M}_d\times S^d$, where $S^d$ is a round $d$-sphere of radius $L$ and metric:
\begin{equation}
    ds^2=g_{\mu\nu}\,dx^\mu\otimes dx^\nu+g_{ij}\,d\upxi^i\otimes d\upxi^j\,,
\end{equation}
where $g_{ij}$ is the metric on the sphere, $\mu,\,\nu,\dots=0,\dots,d-1$ and $i,j,\dots=1,\dots, d$.
We take for ${\bf H}_{(d)}$ the following ansatz:
\begin{equation}
{\bf H}_{(d)}=-\Omega\cdot \boldsymbol{\mathcal{M}}(\phi)\cdot \boldsymbol{\Gamma}\,\boldsymbol{\epsilon}_{{\bf M}_d}+\boldsymbol{\Gamma}\,\boldsymbol{\epsilon}_{{S^d}}\,,
\end{equation}
where
\begin{equation}
\boldsymbol{\epsilon}_{{\bf M}_d}\equiv  \frac{\tilde{e}_{d}}{d!\,L^d}\,\epsilon_{\mu_1\dots \mu_d}\,dx^{\mu_1}\wedge \dots\,dx^{\mu_{d}}\,\,,\,\,\,\boldsymbol{\epsilon}_{{S^d}}\equiv \frac{{e}_{d} }{d!\,L^d}\,\epsilon_{i_1\dots i_d}\,d\upxi^{i_1}\wedge \dots\,d\upxi^{i_{d}}\,,
\end{equation}
where $\tilde{e}_{d}=\sqrt{|{\rm det}(g_{\mu\nu})|}$, ${e}_{d}=\sqrt{{\rm det}(g_{ij})}$. The $(m+n)$-dimensional vector $\boldsymbol{\Gamma}=(\Gamma^M)$ describes the quantized charges:
\begin{equation}
    \Gamma^M=\frac{1}{\mathbb{S}^d}\,\int H^M_{(d)}\in \Lambda^{m,n}\,,
\end{equation}
$\mathbb{S}^d$ being the surface area of a $d$-sphere of radius 1 and $\Lambda^{m,n}$ the even, self-dual charge lattice. Quantum effects break $G$ to its discrete subgroup $G(\mathbb{Z})$ such that $\mathscr{R}[G(\mathbb{Z})]\equiv \mathscr{R}[G]\cap {\rm O}(m,n;\,\mathbb{Z})$, which leaves the charge lattice invariant. The quantum moduli space reads: $G(\mathbb{Z})\backslash G/H.$
Inserting the above ansatz for the $d$-form field strengths, we find, for the scalar fields, the following equation:
$$D_{\hat{\mu}}(\partial^{\hat{\mu}}\phi^s)=\mathscr{G}(\phi)^{st}\,\frac{\partial}{\partial \phi^t}V(\phi;\boldsymbol{\Gamma})\,L^{-2d}\,,$$
where we have introduced the effective scalar potential:
\begin{equation}
V(\phi;\boldsymbol{\Gamma})\equiv \frac{1}{2}\,\boldsymbol{\Gamma}^T\cdot \boldsymbol{\mathcal{M}}(\phi)\cdot \boldsymbol{\Gamma}>0\,.
\end{equation}
Let $\Lambda_1$ denote the smallest sublattice of $\Lambda^{m,n}$ containing $\boldsymbol{\Gamma}$ and $\Lambda_0$ be its complement: $\Lambda^{m,n}=\Lambda_0\oplus \Lambda_1$.
Let $G_1$ denote the largest subgroup of $G$ such that $G_1(\mathbb{Z})$ has a trivial action on $\Lambda_0$ and maps $\Lambda_1$ into itself. Let $G_0$, on the other hand, be a subgroup of $G$ such that $G_0(\mathbb{Z})$ has a trivial action on $\Lambda_1$ and a nontrivial one on $\Lambda_0$. Clearly $[G_0,\,G_1]=0$. The symmetric manifold $G_0/H_0\times G_1/H_1$, $H_0,\,H_1$ being the maximal compact subgroups of $G_0,\,G_1$, respectively, is a totally geodesic submanifold of $G/H$. Let us restrict the scalar fields to $g=(g^k)\in G_1/H_1$ and $\varphi=(\varphi^a)\in G_0/H_0$, setting all the other scalars to zero, and define the coset representative as $L(g,\varphi)=L_1(g)\cdot L_0(\varphi)=L_0(\varphi)\cdot L_1(g)$, where $L_1\in G_1/H_1$ and $L_0\in G_0/H_0$.
Since $G_0$, by definition, is contained in the little group of $\boldsymbol{\Gamma}$, $\mathscr{R}[L_0]\cdot \boldsymbol{\Gamma}=\boldsymbol{\Gamma}$ and thus we can easily verify that:
$$V(\phi;\boldsymbol{\Gamma})=V(g;\boldsymbol{\Gamma})\,,$$ namely $\varphi$ are flat directions of $V$. The extremization of $V$ fixes the $g^k$ scalars only to values $g_*=(g_*^k)$ which only depend on the quantized charges:
$$\left.\frac{\partial V}{\partial g^k}\right\vert_{g=g_*}=0\,\,,\,\,\,g_*=g_*(\boldsymbol{\Gamma})\,.$$
We fix, in the solution, the $g$-scalars to this value: $g=g_*$ and we denote by $V_*$ the extremal value of V: $V_*\equiv \left.V\right\vert_{g=g_*}=V_*(\boldsymbol{\Gamma})>0$.\par
Let us complete the description of the solution by distinguishing two cases:
\begin{itemize}
    \item[i)]{ ${\bf M}_d={\rm AdS}_d$: we take $\varphi^a=\mbox{const.}$. The resulting background is the known ${\rm AdS}_d\times S^d$, where $\varphi^a$ play the role of moduli fields;}
     \item[ii)]{ ${\bf M}_d={\rm AdS}_{d-1}\times S^1$: we take $\varphi^a=\varphi^a(\eta)$, where $\eta \in [0,T]$ is the coordinate along $S^1$, to describe a geodesic on $G_0/H_0$. This manifold, being a symmetric submanifold of a symmetric manifold, is \emph{totally geodesic} and thus the geodesic evolution of $\varphi(\eta)$ defines a geodesic curve on $G/H$ and will not source scalar fields which were set to zero. }
\end{itemize}
\paragraph{Case i):}
The general form of the metric is:
$$ds^2=v_1^2\,ds^2_{{\rm AdS}_d}+ L^2\,ds^2_{S^d}\,,$$
where $ds^2_{{\rm AdS}_d}$ and $ds^2_{S^d}$ are the metrics on ${\rm AdS}_d$ and $S^d$, with unit radii, respectively. The radius $v_1$ of ${\rm AdS}_d$  and the radius $L$ of the sphere are then fixed by the field equations to the values:
$$v_1=L=\left[\frac{V_*}{2(d-1)}\right]^{\frac{1}{2(d-1)}}\,.$$
\paragraph{Case ii):}
The metric has the form:
\begin{equation}ds^2=v_1^2\,ds^2_{{\rm AdS}_{d-1}}+v_2^2\,d\eta^2+ L^2\,ds^2_{S^d}\,.\label{metric}\end{equation}
The field equations now imply
\begin{equation}
    v_1=\sqrt{\frac{d-2}{d-1}}\,L\,,\,\,\,v_2=\,\frac{\kappa}{\sqrt{d-1}}\,L\,,\,\,\,L=\left[\frac{V_*}{2(d-1)}\right]^{\frac{1}{2(d-1)}}\,,
\end{equation}
where $\kappa$ is the "velocity" of the geodesic on $G_0/H_0$:
$$\kappa^2=\frac{1}{2}\,\mathscr{G}_{ab}(\varphi)\,\dot{\varphi}^a\dot{\varphi}^b\,,\,\,\,\dot{\varphi}^a\equiv \frac{d\varphi^a}{d\eta}\,.$$
This can be verified by using, in the field equations, the following explicit form of the curvature components:
\begin{align}
    R_{\alpha\beta\gamma\delta}&= -v_1^{-2}\,(g_{\alpha \gamma} g_{\beta \delta}-g_{\alpha \delta} g_{\beta \gamma})\,\,\Rightarrow\,\,\,R_{\alpha\beta}=R_{\alpha\gamma\beta}{}^\gamma=-\frac{(d-2)}{v_1^2}\,g_{\alpha\beta}\,\nonumber\\
   R_{ijkl}&= L^{-2}\,(g_{ik} g_{jl}-g_{il} g_{jk})\,\,\Rightarrow\,\,\,R_{ij}=R_{ikj}{}^k=\frac{(d-1)}{L^2}\,g_{ij}\nonumber\\
   R_{\eta\eta}&=0\,,
\end{align}
where $\alpha,\beta,\dots=0,\dots,d-2 $ label the coordinates on ${\rm AdS}_{d-1}$, and of the non-vanishing components of $T^{(H)}$:
\begin{align} T^{(H)}_{\alpha\beta}&=-\frac{1}{2}\,V_*\,L^{-2d}\,g_{\alpha\beta}\,,\nonumber\\
T^{(H)}_{\eta\eta}&=-\frac{1}{2}\,V_*\,v_2^2\,L^{-2d}\,,\nonumber\\
T^{(H)}_{ij}&=\frac{1}{2}\,V_*\,L^{-2d}\,g_{ij}\,.\nonumber
\end{align}
\paragraph{$U$-Fold structure of the solution $ii)$.} The ${\rm AdS}_{d-1}\times S^1\times S^d$ background described above can be a consistent solution to the underlying quantum theory provided the ending points $P_0\equiv (\varphi^a(0))$ and $P_1\equiv (\varphi^a(T))$ of the geodesic curve in $G_0/H_0$ are identified in the quantum moduli space, namely if
$$\exists \mathfrak{M}\in G_0(\mathbb{Z})\,:\,\,\,\mathfrak{M}\cdot P_0=P_1\,.$$
Such solutions are defined by conjugacy classes of $\mathfrak{M}$ in $G_0(\mathbb{Z})$.\par
A geodesic on $G_0/H_0$ with initial point $P_0=\varphi(0)$ and "velocity vector" $\mathbb{Q}$ is solution to the matrix equation:
\begin{equation}
\mathcal{M}_0(\varphi(\eta))=\mathcal{M}_0(\varphi(0))\cdot e^{\mathbb{Q}^T\eta}=L_0(\varphi)\cdot  e^{\mathbb{Q}_0\eta}\cdot L_0(\varphi)^T\,,
\end{equation}
where $\mathcal{M}_0(\varphi)\equiv L_0(\varphi)\cdot L_0(\varphi)^T$, $\mathbb{Q}\in T_{\varphi(0)}(G_0/H_0)$ and $\mathbb{Q}_0=\mathbb{Q}_0^T\in T_{O}(G_0/H_0)$, $O$ being the origin of $G_0/H_0$ where $\varphi^a=0$. Given an initial point $P_0$ and a monodromy matrix $\mathfrak{M}\in G_0(\mathbb{Z})$, the corresponding geodesic is defined by solving the equation:
$${\bf L}(\varphi(0))\cdot e^{\mathbb{Q}_0 T}\cdot {\bf L}(\varphi(0))^T=\mathfrak{M}\cdot{\bf L}(\varphi(0))\cdot {\bf L}(\varphi(0))^T\cdot \mathfrak{M}^T\,,$$
in $\mathbb{Q}_0 T$, in the appropriate matrix representation.

\section{Application of the General Construction}
Let us discuss some explicit constructions of $U$-folds, applying the general procedure discussed above.
\paragraph{Type IIB in $2d=10$ dimensions.} We consider Type IIB superstring theory and its low-energy effective supergravity description, which is a chiral maximal ten-dimensional supergravity  \cite{Schwarz:1983qr}. The latter, at the classical level, features a global symmetry group $G={\rm SL}(2,\mathbb{R})_{{\rm IIB}}$, which contains the quantum global symmetry group  $G(\mathbb{Z})={\rm SL}(2,\mathbb{Z})_{{\rm IIB}}$ of the corresponding superstring theory. The bosonic sector consists, aside from the metric, of an axion-dilaton complex field $\rho=C_{(0)}+i\,e^{-\Phi}$ spanning the manifold ${\rm SL}(2,\mathbb{R})_{{\rm IIB}}/{\rm SO}(2)$, an ${\rm SL}(2,\mathbb{R})_{{\rm IIB}}$-doublet of 2-forms $B^\upalpha_{(2)}=(B_{(2)},\,C_{(2)})$, and a R-R 4-form $C_{(4)}$, whose field strength $\hat{F}_{(5)}$ is self-dual: $\hat{F}_{(5)}={}^*\hat{F}_{(5)}$. This property implies that the representation $\mathscr{R}$ of $G$, acting on the 5-form field strength and its dual, is trivial: $\mathscr{R}=1$. The ${\rm AdS}_5\times S^5$ background is well known \cite{Schwarz:1983qr} and describes the near-horizon geometry of a stack of overlapping $D3$-branes.\par
The  ${\rm AdS}_4\times S^1\times S^5$ J-fold solution was constructed in \cite{Guarino:2019oct} and features vanishing 2-form fields $B^\upalpha_{(2)}=0$. Choosing as monodromy $\mathfrak{M}=J_{\tt n}\in {\rm SL}(2,\mathbb{Z})_{{\rm IIB}}$, for ${\tt n}>0$, the corresponding geodesic in the axion-dilaton fields reads:
\begin{align}
    e^{\Phi(\eta)}&=\frac{{\tt n} \sinh \left(\sqrt{2}\kappa \eta\right)}{\sqrt{{\tt n}^2+4}}+\cosh \left(\sqrt{2}\kappa \eta\right)\,\,,\,\,
   C_{(0)}(\eta)=-\frac{2 \sinh \left(\sqrt{2}\kappa \eta\right)}{\sqrt{{\tt n}^2+4} \cosh \left(\sqrt{2}\kappa \eta\right)+{\tt n}\sinh \left(\sqrt{2}\kappa \eta\right)}\nonumber\\
   T&=\frac{1}{\sqrt{2} \kappa }\cosh ^{-1}\left(\frac{{\tt n}^2}{2}+1 \right)\,.
\end{align}
As we move around $S^1$, the axion-dilaton field undergoes the non-perturbative transformation: $\rho(0)\rightarrow \rho(T)=-\frac{1}{\rho(0)+{\tt n}}$.
\paragraph{Type IIB on $CY_2$ ($d=3$).}  We shall consider the cases in which $CY_2=T^4$, $CY_2=K_3$ \cite{Aspinwall:1994rg} and, finally, $CY_2=T^4/\mathbb{Z}_2$ in the presence of an orientifold 5-plane $\mathcal{O}_5$ \cite{Polchinski:1998rr,Angelantonj:2002ct}.\par
In the first two cases, $G={\rm O}(5,n)$ where $n=5$ for $CY_2=T^4$ and $n=21$ for $CY_2=K_3$. The scalar manifold is $\mathscr{M}_{\rm scal}={\rm O}(5,n)/{\rm O}(5)\times {\rm O}(n)$ and the resulting six-dimensional supergravities are the maximal $\mathcal{N}=(2,2)$ and the $\mathcal{N}=(2,0)$, respectively (see \cite{Sezgin:2023hkc} for a survey of supergravities in diverse dimensions, and references therein). The tensor fields $B^M_{(2)}$ are $5+n$, in the fundamental representation of $G$. The former model also features 16 vector fields in the spinorial representation of ${\rm Spin}(5,n)$ which, however, we set to zero in our solutions.
As far as the compactification on a $T^4/\mathbb{Z}_2$-orientifold is concerned, the resulting theory is $D=6$ $\mathcal{N}=(1,1)$ with a single tensor field and 8 vector fields \cite{Dibitetto:2019odu}. In this case $G={\rm SO}(1,1)\times {\rm SO}(4,4)$, where only the ${\rm SO}(1,1)$ acts, through the 2-dimensional representation $\mathscr{R}$, on the tensor field strength and its dual. We shall expand on this model in the next Section.
The solution ${\rm AdS}_3\times S^3\times CY_2$ describes the near-horizon geometry of a D1-D5 system or of duality-related ones, such as the F1-NS5 system, where the 5-branes, in the directions transverse to the 1-branes, wrap $CY_2$ \cite{Seiberg:1999xz}. The D1-D5 charges $Q_1,\,Q_5$ span a $\Lambda_1=\Lambda^{1,1}$ lattice.\footnote{In the $CY_2=T^4$ case, the charge parameters $Q_1,\,Q_5$ are often denoted in the literature \cite{Polchinski:1998rr} by $r_1^2,\,r_5^2$, respectively, and enter the harmonic functions $Z_1,\,Z_5$ describing the D1-D5 solution as $Z_u=1+\frac{Q_u}{r^2}$, $u=1,5$. They are related to the corresponding quantized string fluxes ${\bf Q}_1,{\bf Q}_5$ as follows:
$$Q_1=g_s\,{\bf Q}_1 (\alpha')^{\mbox{\tiny 3}}\,\frac{(2\pi)^4}{V_4}\,,\,\,\,Q_5=g_s\,{\bf Q}_5 \alpha'\,,$$
where $V_4/(2\pi)^4\equiv ({\rm det}(G_{(0){\, ab}}^{(s)}))^\frac{1}{2}$, where $G_{{(0)\, ab}}^{(s)},\,G_{{(0)\, ab}}$  are the metric on $T^4$ in the string and Einstein frames at radial infinity in the D1-D5 solution, respectively. We can fix the $D=6$ Newton's constant and $V_4$ so that $Q_1={\bf Q}_1$ and $Q_5={\bf Q}_5$.} In the $CY_2=T^4$ or $K_3$, cases, the 2-form charge lattice splits as $\Lambda^{5,n}=\Lambda^{1,1}\oplus \Lambda^{4,n-1}$, where $\Lambda_2=\Lambda^{4,n-1}$ is the complement to $\Lambda^{1,1}$, and is acted on by ${\rm O}(4,n-1;\mathbb{Z})\subset G(\mathbb{Z})$. In the orientifold reduction, on the other hand, the 2-form charge lattice is only $\Lambda^{1,1}$. In all cases $G_1={\rm O}(1,1)$. In defining our new solutions, according to our previous analysis, we consider the submanifold:
$$\frac{G_1}{H_1}\times \frac{G_0}{H_0} ={\rm O}(1,1)\times \frac{G_0}{H_0}\subset {\rm O}(1,1)\times \frac{{\rm O}(4,n-1)}{{\rm O}(4)\times {\rm O}(n-1)}\subseteq \mathscr{M}_{{\rm scal.}}\,,$$
where $n=21$ only for $CY_2=K_3$ while $n=5$ in all the other cases. The monodromy of the $U$-fold solution is a matrix of $\mathfrak{M}$ in the group $G_0(\mathbb{Z})\subset {\rm O}(4,n-1;\mathbb{Z})$.\par
For the sake of concreteness, let us focus on the case $CY_2=T^4$ (or the $T^4/\mathbb{Z}_2$-orientifold).
The factor $\frac{G_1}{H_1}={\rm O}(1,1)$ is now parametrized by a single scalar $g$ defined as $e^g\equiv e^\Phi\,{\rm det}(G_{{ ab}})^{\frac{1}{2}}$, $G_{{ab}}$ being the metric on $T^4$ in the Einstein frame, while the submanifold $\frac{{\rm O}(4,4)}{{\rm O}(4)\times {\rm O}(4)}$ is spanned by the moduli $\tilde{G}_{{ ab}}\equiv e^{-\frac{\Phi}{2}}\,{G}_{{ab}}$ and $C_{{ ab}}$, internal components of the ten-dimensional R-R 2-form field. Writing the charge vector according to the above splitting of the lattice $\Gamma^M=2(Q_5,\,Q_1;\,0,\dots,0)$ (or $\Gamma^M=2(Q_5,\,Q_1)$ in the orientifold case), the potential reads $$V=2(Q_1^2\,e^{-g}+Q_5^2\, e^g)\,\,\Rightarrow\,\,\,\,e^{g_*}=\frac{Q_1}{Q_5}\,\,,\,\,\,V_*=4Q_1 Q_5\,.$$
The metric has the form \eqref{metric} with $d=3$, namely
\begin{equation}
ds^2=L^2\,\left(\frac{1}{2}\,ds^2_{{\rm AdS}_2}+\frac{\kappa^2}{2}\,d\eta^2+ds^2_{S^3}\right)\,\,,\,\,\,L=(Q_1Q_5)^{\frac{1}{4}}\,. \label{6D metric}
\end{equation}
To give an explicit solution for the moduli fields, we can restrict the latter to the only fields \cite{Astesiano:2022qba}:
\begin{align}
    (\varphi^a)&=(\rho_1,\,\rho_2)=(C_{12}+i\,\tilde{G}_{11},C_{34}+i\,\tilde{G}_{33})\in  \frac{G_{\mbox{\tiny 0}}}{H_{\mbox{\tiny 0}}}=\left(\frac{{\rm SL}(2,\mathbb{R})}{{\rm SO}(2)} \right)^2\subset
\frac{{\rm O}(4,4)}{{\rm O}(4)\times {\rm O}(4)}\,,\nonumber\\
\tilde{G}_{22}&=\tilde{G}_{11}\,,\,\,\,\tilde{G}_{44}=\tilde{G}_{33}\,.
\end{align}
As monodromy matrix, we can choose:
\begin{equation}
    \mathfrak{M}=\mathfrak{M}_{\mbox{\tiny 1}}\cdot\mathfrak{M}_{\mbox{\tiny 2}}=J_{\mbox{\tiny ${\tt n}_1$}}\cdot J_{\mbox{\tiny ${\tt n}_2$}}\in {\rm SL}(2,\mathbb{Z})\times{\rm SL}(2,\mathbb{Z})\subset
{{\rm O}(4,4;\,\mathbb{Z})}\,\,;\,\,\,{\tt n}_1,\,{\tt n}_2>0\,.
\end{equation}
The explicit geodesic solution, in term of the constants $\tt{c}_{1}=\cosh^{-1}(\frac{1}{2}{\tt n}_{1}^{2}+1)$ and $\tt{c}_{2}=\cosh^{-1}\left(\frac{1}{2}{\tt n}_{2}^{2}+1\right)$, reads \cite{Astesiano:2024gzy}:
\begin{align}
\tilde{G}_{22}(\eta)^{-1} & =\tilde{G}_{11}(\eta)^{-1}=\cosh({\textstyle \frac{\mathtt{c}_{1}}{T}\eta})+\frac{{\tt n}_{1}}{\sqrt{{\tt n}_{1}^{2}+4}}\sinh({\textstyle \frac{\mathtt{c}_{1}}{T}\eta})\,,\nonumber \\
\tilde{G}_{44}(\eta)^{-1} & =\tilde{G}_{33}(\eta)^{-1}=\cosh({\textstyle \frac{\mathtt{c}_{2}}{T}\eta})+\frac{{\tt n}_{2}}{\sqrt{{\tt n}_{2}^{2}+4}}\sinh({\textstyle \frac{\mathtt{c}_{2}}{T}\eta})\,,\nonumber \\
C_{12}(\eta) & =-\frac{2 }{\sqrt{{\tt n}_{1}^{2}+4}\coth({\textstyle \frac{\mathtt{c}_{1}}{T}\eta}) + {\tt n}_{1} }\,,\hspace{1cm}
C_{34}(\eta)=-\frac{2}{\sqrt{{\tt n}_{2}^{2}+4}\coth({\textstyle \frac{\mathtt{c}_{2}}{T}\eta})+{\tt n}_2 }\,, \nonumber \\
\kappa^{2} & =\frac{\mathtt{c}_{1}+\mathtt{c}_{2}}{2T^{2}}\,.
\end{align}
As $\eta:\,0\,\rightarrow\,T$ the two complex moduli vary accordingly $\rho_\ell(0)\,\rightarrow\,\rho_\ell(T)=-\frac{1}{\rho_\ell+{\tt n}_\ell}$, $\ell=1,2$.\par The ten-dimensional string coupling constant, in particular, by effect of the monodromy, undergoes the following variation:
$$
e^{\Phi(0)}=\left(\frac{Q_{\mbox{\tiny 1}}}{Q_{\mbox{\tiny 5}}}\right)^{\frac{1}{2}}\,\longrightarrow\,\,\,e^{\Phi(T)}=\left(\frac{Q_{\mbox{\tiny 1}}}{Q_{\mbox{\tiny 5}}}\right)^{\frac{1}{2}}\,\sqrt{({\tt n}_1+1)({\tt n}_2+1)}\,.
$$
In order to assess the supersymmetry of the solution, we focus on a special case within the $D=6$ model originating from the compactification on the $T^4/\mathbb{Z}_2$-orientifold. We discuss this in the last section and prove that the $U$-fold solution under consideration is, just as its counterpart in \cite{Guarino:2019oct}, non-supersymmetric. This analysis is new and was not contained in \cite{Astesiano:2024gzy}.
\section{$U$-Folds in the Orientifold Model and Supersymmetry}
We discuss here the new solutions within the half-maximal $\mathcal{N}=(1,1)$, $D=6$, ungauged supergravity originating from the compactification of Type IIB superstring theory on a $T^4/\mathbb{Z}_2$-orientifold.
The resulting theory describes the invariant sector with respect to the action of the involution $\Omega \mathcal{I}_4$, where $\Omega$ denotes the worldsheet parity and $\mathcal{I}_4$ the parity along the directions of the torus: $y^a\rightarrow -y^a$, $y^a$ being the coordinates on $T^4$.
The bosonic sector of the reduced theory consists, aside from the metric, of the dilaton $\Phi$, the internal metric $G_{ab}$, the internal components $C_{ab}$ of the R-R 2-form field $C_{(2)}$, four vectors $B_{(1)\,a}$ originating from the reduction of the Kalb-Ramond field and four vectors $C^a_{(1)}\equiv \epsilon^{abcd}\,C_{(1)\,bcd}/3!$ from the R-R 4-form. Finally, the six-dimensional supergravity features a single 2-form $C_{(2)}$ from the 10-dimensional R-R 2-form field. The scalar fields span the manifold
 \begin{equation}
   \mathscr{M}_{{\rm scal.}}=\frac{G}{H}= {\rm O}(1,1)\times \frac{{\rm O}(4,4)}{{\rm O}(4)\times {\rm O}(4)}\,,
 \end{equation}
 where ${\rm O}(1,1)$ is parametrized by the scalar $g$ defined above and the $\frac{{\rm O}(4,4)}{{\rm O}(4)\times {\rm O}(4)}$
 factor is spanned by $\tilde{G}_{ab},\,C_{ab}$, introduced earlier. The eight vectors $(C^a_{(1)},\,B_{(1)\,a})$ transform in the fundamental representation of ${\rm O}(4,4)$ while only the ${\rm O}(1,1)$ factor in $G$ acts non-trivially on the 3-form tensor field strength and its dual ($\mathscr{R}$ is not faithful).\par After computing the solution in this $D=6$ model, we uplifted it to the $D=10$ theory. In the $U$-fold construction the monodromy matrix $\mathfrak{M}$ belongs to ${\rm O}(4,4;\,\mathbb{Z})$. For the sake of assessing supersymmetry, we shall, however, consider the simplest case in which the only non-vanishing modulus is related to the volume of the internal torus by writing:
 $$\tilde{G}_{ab}=e^{2\sigma(\eta)}\,\delta_{ab}\,.$$
 Since $\mathscr{G}_{\sigma\sigma}=8$, the geodesic is defined by
 $$\sigma(\eta)=-\frac{\kappa}{2}\,\eta\,.$$
 Note that this geodesic does not allow for a non-trivial
integer monodromy matrix. To have a possible consistent solution, we would still need to act on this background with the global symmetry of the classical theory ${\rm O}(4,4)$ and fix $T$ appropriately. Since the last procedure will not affect supersymmetry, we shall focus on the 1-modulus simplest version of it described above to assess this latter property of this background.

Recall that the scalar $g=e^{2\Phi}\,({\det \tilde{G}_{ab}})^\frac{1}{2}$ is fixed to $g_*=Q_1/Q_5$, which implies that the ten dimensional dilaton reads:
 $$e^{\Phi}=\sqrt{\frac{Q_1}{Q_5}}\,e^{\kappa \eta}\,.$$
 The 3-form field strength has the form:
$$F_{3}=2Q_{1}e^{-g_*}\boldsymbol{\epsilon}_{{\rm AdS}_2\times S^1}+2Q_{5}\boldsymbol{\epsilon}_{S^{3}}=2 Q_5\left(\boldsymbol{\epsilon}_{{\rm AdS}_2\times S^1}+\boldsymbol{\epsilon}_{S^{3}}\right)\,,
$$
where
$$\boldsymbol{\epsilon}_{{\rm AdS}_2\times S^1}=\frac{\kappa}{2\sqrt{2}}\,\boldsymbol{\epsilon}_{{\rm AdS}_2}\wedge d\eta\,,$$
and $\boldsymbol{\epsilon}_{{\rm AdS}_2}$ is the volume form of ${\rm AdS}_2$ with unit radius.
Finally, the ten-dimensional metric in the Einstein frame, obtained by uplifting the $D=6$ solution, reads:
\begin{align*}
ds_{10}^{2} & =\left(\frac{Q_{5}}{Q_{1}}\right)^{\frac{1}{4}}e^{\frac{\kappa\eta}{2}}L^{2}\left(\frac{1}{2}ds_{\mathrm{AdS}_{2}}^{2}+\frac{\kappa^{2}}{2}d\eta^{2}+ds_{S^{3}}^{2}\right)+\left(\frac{Q_{1}}{Q_{5}}\right)^{\frac{1}{4}}e^{-\frac{\kappa\eta}{2}}dy^{a}\otimes dy^{a}\,.
\end{align*}
We then evaluated the variation of the ten-dimensional spin-$1/2$ fields of Type IIB supergravity on the uplifted solution:\footnote{We suppress the R-symmetry indices of the $D=10$ spinors. The Pauli matrices $\sigma^x$, $x=1,2,3$, are meant to act on these indices. }
\begin{align}
\delta\lambda\propto\left(\not{d}g+\frac{1}{2}\not{d}\log\hat{G}+e^{-\frac{g}{4}}\Gamma_{*}\not H_{2a}\hat{E}_{\underline{a}}{}^{a}\Gamma^{\underline{a}}\sigma_{3}-e^{\frac{g}{2}}\not F_{3}\sigma_{1}-\frac{1}{2}\hat{G}^{\frac{1}{2}}\not{d}C_{ab}\hat{E}_{\underline{a}}{}^{a}\hat{E}_{\underline{b}}{}^{b}\Gamma^{\underline{ab}}\sigma_{1}\right)\epsilon\,, \label{var lamb 6d}
\end{align}
where we defined $H_{2a}\equiv dB_{(1)\,a}$, $\hat{G}_{ab}\equiv ({\rm
 det}(\tilde{G}_{ab}))^{-\frac{1}{2}}\,\tilde{G}_{ab}$, $\hat E_a{}^{\underline{a}}$ denotes the corresponding vielbein and $\hat{G}\equiv\det(\hat{G}_{ab})$. The matrices $\Gamma^{\underline{a}}$
close the 4-dimensional Euclidean Clifford algebra. Given a $p$-form
in 6 dimensions $\zeta$, we define $\not{\zeta}=\frac{1}{p!}\zeta_{\underline{\mu_{1}\dots\mu_{p}}}\gamma^{\underline{\mu_{1}\dots\mu_{p}}}$,
where $\gamma^{\underline{\mu}}$ are 6-dimensional Lorentzian
 gamma-matrices and rigid indices are underlined. The
consistent orientifold truncation requires the following projectors to hold $\Gamma_{*}\sigma_{1}\lambda=\pm\lambda,$
$\Gamma_{*}\sigma_{1}\epsilon=\mp\epsilon$ together with the chirality
constraint coming from type IIB supergravity $\gamma_{*}\Gamma_{*}\lambda=-\lambda$
and $\gamma_{*}\Gamma_{*}\epsilon=\epsilon$. The matrices $\gamma_{*},\Gamma_{*}$
are the chirality matrices in 6 and 4 dimensions, respectively. \par
We find that, on this background, $\delta \boldsymbol{\lambda}\neq 0$, indicating that no supersymetry is preserved.
\section{Conclusions}
We have reviewed a general construction of $U$-fold solutions to Type IIB superstring theory, with geometry ${\rm AdS}_{d-1} \times S^1 \times S^d\times M_{10-2d}$ and monodromy along $S^1$ in the U-duality group $G(\mathbb{Z})$ of the ten-dimensional theory, compactified on the Ricci-flat manifold $M_{10-2d}$. Explicit examples of these solutions are constructed in Type IIB supergravity, in the case $d=3$. These solutions are characterized by a geodesic evolution of the moduli fields $\varphi^a(\eta)$ in the moduli space $G_0/H_0$. Such dependence of the fields on the $\eta$ coordinate along $S^1$ is equivalently described by a Cremmer-Scherk-Schwarz (CSS) reduction on $S^1$, with non-compact twist $\mathcal{A}(\eta)=L_0(\varphi^a(\eta))\in G_0$, of the $2d$-dimensional theory originating from the reduction of Type IIB supergravity on $M_{10-2d}$. The resulting theory in $(2d-1)$-dimensions features a background of the form ${\rm AdS}_{d-1} \times S^d$. Although the prototypical solutions discussed here are non-supersymmetric, they capture some distinctive features of the most general background of this kind.\par
A natural sequel of this analysis would be to construct supersymmetric $U$-fold solutions with geometry ${\rm AdS}_2\times S^1\times \tilde{S}^3$, generalizing those discussed here, featuring a deformation $\tilde{S}^3$ of ${S}^3$ and possibly vector fields in the six dimensional supergravity theory. These backgrounds would correspond to the supersymmetric ${\rm AdS}_4\times S^1\times \tilde{S}^5$ $J$-folds constructed in \cite{Inverso:2016eet}, where the role of the six-dimensional vector fields is played by the ten-dimensional 2-forms.\par
Recently, in \cite{Guarino:2024zgq}, supersymmetric ${\rm AdS}_2$ $J$-folds, with geometry ${\rm AdS}_2\times \Sigma_2\times S^1\times \tilde{S}^5$ were constructed in Type IIB supergravity, $\Sigma_2$ being a compact Riemannian surface of genus $g>1$. A subclass of these describe the near-horizon geometries of BPS universal black holes asymptoting the ${\rm AdS}_4\times S^1\times \tilde{S}^5$ $\mathcal{N}=4$ $J$-fold, or its $\mathcal{N}=2$ marginal deformations. \par
 Supersymetric ${\rm AdS}_2$ $U$-folds could, in principle, also be constructed by compactifying the Janus solutions constructed in \cite{Chiodaroli:2009yw}. The holographic dual of these ${\rm AdS}_2\times S^1\times \tilde{S}^3\times CY_2$ $U$-folds, in analogy with their ten-dimensional counterparts, is naturally identified with the IR limit of the interface $(1+0)$-theory within the $1+1$ SCFT dual to the corresponding D1-D5 system\footnote{This theory is conjectured to be described, in an appropriate limit, by the hyper-K\"ahler sigma model whose target space is symmetric product orbifold ${\rm Sym}^N(CY_2)$ \cite{Seiberg:1999xz}.}, with a monodromy acting on the moduli along $S^1$.\par
 Partial results are obtained by performing a CSS compactification to $D=2$ from half-maximal $D=3$ gauged supergravity featuring supersymmetric $AdS_3$ vacua, with CSS-twist matrix in the isometry group of the moduli space of the $AdS_3$ vacua \cite{wip}.

\section{Acknowledgements}
M.T. is deeply indebted to Davide Astesiano and Daniele Ruggeri, with whom the results of \cite{Astesiano:2024gzy}, partly reviewed here, were obtained. The authors would also like to thank Henning Samtleben for interesting discussions. R.N. was supported by the European Union and the Czech Ministry of Education, Youth and
Sports (Project: MSCA Fellowship CZ FZU III - CZ.02.01.01/00/22$\_$010/0008598). M.O. is partially supported by Beca ANID de Doctorado grant 21222264.

\bibliographystyle{JHEP}

\end{document}